\documentclass[%
 reprint,
 amsmath,amssymb,
 aps,
]{revtex4-1}

\usepackage{graphicx}
\usepackage{dcolumn}
\usepackage{bm}

\begin{document}

\preprint{APS/123-QED}

\title{Link-based formalism for time evolution of adaptive networks}

\author{Jie Zhou}
\email{jzhou@phy.ecnu.edu.cn}

\affiliation{%
Institute of Theoretical Physics and Department of Physics,
East China Normal University, Shanghai, 200062, China.}%

\affiliation{%
Department of Electronic Engineering, City University of Hong Kong, Hong Kong SAR, China.}%

\author{Gaoxi Xiao}%
\email{EGXXiao@ntu.edu.sg}
\affiliation{%
Division of Communication Engineering, School of Electrical and
Electronic Engineering, Nanyang Technological University,
Singapore 639798.
}%

\affiliation{%
Complexity Program, President's Office, Nanyang Technological University, Singapore 639673.}%

\author{Guanrong Chen}
\affiliation{%
Department of Electronic Engineering, City University of Hong Kong, Hong Kong SAR, China.}%


\begin{abstract}
Network topology and nodal dynamics are two fundamental stones of 
adaptive networks. Detailed and accurate knowledge of these two 
ingredients is crucial for understanding the evolution and 
mechanism of adaptive networks. In this paper, by adopting the 
framework of adaptive SIS model proposed by Gross \emph{et al.} 
[Phys. Rev. Lett. \textbf{96}, 208701 (2006)] and carefully 
utilizing the information of degree correlation of the network, 
we propose a link-based formalism for describing the system 
dynamics with high accuracy and subtle details. Several specific 
degree correlation measures are introduced to reveal the 
co-evolution of network topology and system dynamics.
\begin{description}
\item[PACS numbers] 89.75.Hc, 87.19.X-, 89.75.Fb.
\end{description}
\end{abstract}

\pacs{Valid PACS appear here}
\maketitle


\section{Introduction}
\label{sec:Introduction}

Binary-state dynamics on complex networks have been widely 
adopted to model various social and technological systems. Such 
modelling is highly competent in system description while still 
allowing relatively simple analysis. In a network with 
binary-state dynamics, nodes can switch between two discrete 
states, and the rules of switching typically depend on the nodal 
dynamics and the network structure. Examples include information 
cascade dynamics \cite{Duncan:2002}, opinion dynamics 
\cite{Nardini:2008,Holme:2006,Vazquez:2008}, disease propagation 
\cite{Satorras:2001,Barthelemy:2004,Noel:2009,Zhou:2012b,Liu:2005,Tang:2009}, 
rumor spreading \cite{Zhou:2007}, Ising model 
\cite{Dorogovtsev:2002}, and neural dynamics 
\cite{Goltsev:2010,Acebron:2007,Zhou:2011}.

Among these models, adaptive epidemic
susceptible-infected-susceptible (SIS) model \cite{Gross:2006} 
has been commonly used to study adaptive networks with 
co-evolution of epidemic dynamics and network topology 
\cite{Gross:2006,Zhou:2012a}. Efforts have been made to construct 
a theoretical foundation for predicting the time evolution of 
such a system and its dynamical features 
\cite{Gross:2008,Marceau:2010} and several successful node-based 
methods have been developed. For example, by adopting a node 
classification approach, where nodes are distinguished by their 
states and the states of their neighbors, a highly accurate 
mean-field theory-based approach has been proposed to predict the 
evolutions of adaptive networks \cite{Marceau:2010}. A 
theoretical framework has also been built to describe general 
binary-state dynamics with high-order accuracy 
\cite{Gleeson:2011,Demirel:2012,Durrett:2012,Gleeson:2013}. 
However, all the current node-based methods employ \emph{zero 
degree correlation} approximation, in which the degree 
correlation of the underlying networks is largely ignored. Such 
an approximation imposes limitations to revealing detailed 
network structures, causing nontrivial degradation in the 
accuracy of prediction, especially in correlated networks. A 
theoretical formalism that can properly reveal detailed network 
structural information is therefore in great demand.

In this work, based on the framework of the adaptive epidemic 
model of Gross \emph{et. al.} \cite{Gross:2006}, we introduce a 
theoretical formalism from a link classification approach. 
Differing from the node classification method, where network 
nodes are the objects of classification, in the proposed link 
classification approach links are the objects to be classified 
according to the infection states, the number of neighboring 
links, and the number of infected neighboring nodes of each 
link's two end nodes, respectively. As we shall show, compared to 
the existing results, the new approach achieves more accurate 
prediction of the time evolution of adaptive networks and 
provides more detailed information on the network structure, 
thanks to a proper reflection of degree correlation.

The remainder of this paper is organized as follows. We introduce
the adaptive network model In Sec. \ref{sec:Model} and present 
our formalism in Sec. \ref{sec:Formalism}. We then validate the 
proposed formalism by simulation results for uncorrelated and 
correlated network, respectively, in Sec. \ref{sec:Results}. 
Finally, We conclude the paper by Sec. \ref{sec:Conclusion}.

\section{Adaptive epidemic network Model}
\label{sec:Model}

For clarity of later discussions, we first introduce some common 
definitions about the network structure under investigation. 
Suppose that there are $N$ nodes and $M$ edges in a network, the 
average nodal degree $\langle k \rangle$ therefore equals $2M/N$. 
The degree distribution of the network is denoted by $p_k$, which 
represents the probability that the degree of a randomly selected 
node is $k$. The \emph{edge end distribution} is denoted as 
$q_k$, which represents the probability that the degree of a node 
reached by a randomly selected edge is $k$. Both $p_k$ and $q_k$ 
satisfy normalization conditions, i.e., $\sum_k{p_k=1}$ and 
$\sum_k{q_k=1}$. Thus, $\langle k \rangle=\sum_k{kp_k}$ and 
$q_k=kp_k/\langle k \rangle$ \cite{Newman:2002,Weber:2007}.

The degree correlation of a network can be represented by a joint 
degree distribution $p_{i,k}$. Specifically, $p_{i,k}$ denotes 
the probability that the two end nodes of a randomly selected 
edge have degrees $i$ and $k$, respectively. For an undirected 
network, $p_{i,k}=p_{k,i}$. According to the definition of 
$p_{i,k}$, it is straightforward to have $q_k=\sum_i p_{i,k}$ and 
$kp_k=\langle k \rangle\sum_i{p_{i,k}}$ \cite{Weber:2007}. The 
overall degree correlation is described by the 
\emph{assortativity coefficient} $r$ defined as follows 
\cite{Newman:2002}:
\begin{equation}
\label{eq:r} r=\frac{1}{\sigma_e^2}\sum_{i,k}ik[p_{i,k}-q_iq_k],
\end{equation}
where $\sigma_e^2=\sum_k k^2q_k-\left(\sum_k q_k\right)^2$ is the 
normalization factor to make $r\in [-1,1]$. In the global scale, 
$r>0$ ($r<0$) corresponds to positive (negative) correlation, 
where a large-degree node tends to connect another large-degree 
(small-degree) node and vice versa; and $r=0$ means there is not 
such correlation. Note that in the original definition proposed 
in Ref. \cite{Newman:2002}, the calculated of $r$ adopted the 
\emph{remaining} degree of each node, which is its actual degree 
minus one. In our calculations hereafter, we directly adopt the 
actual nodal degree: it is easy to prove that these two 
definitions are equivalent in the limit of large networks.

Next, we briefly introduce the adaptive 
susceptible-infected-susceptible (SIS) model. In the SIS model, 
each node can either be susceptible (S) or infected (I), which 
are the two states of the infection. When an S node is connected 
with an I node, it has a probability of $\beta$ in each time step 
to be infected and becomes an I node. If an S node has a total of 
$k$ neighbors of which $k_\textrm{inf}$ are I nodes, the 
probability that an S node to be infected in a single time step 
equals $1-(1-\beta)^{k_\textrm{inf}}$. When the interval of a 
time step is set to be small enough, $\beta$ shall be a small 
value; hence the probability that an S node with $k_\textrm{inf}$ 
infected neighbors is converted into an I node can be 
approximated by $\beta k_\textrm{inf}$. On the other hand, in 
each time step, an I node has a probability of $\gamma$ to 
recover to S state. In the adaptive network model, we assume that 
all the nodes have global knowledge of node infection state. To 
avoid being infected, an S node may cut each of its links 
connected to infected neighbors with a probability $\omega$. To 
compensate each lost connection the S node will form a new 
connection with a randomly selected S node (not one of its 
existing neighbors). By doing so, the link rewiring process 
remains the total number of links in the network.

\section{Formalism based on link classification}
\label{sec:Formalism}

Now, we introduce the link classification method and derive a 
theoretical formalism accordingly. 

\begin{figure}[b]
\includegraphics[width=0.7\linewidth]{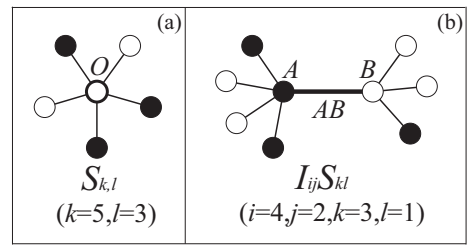}
\caption{\label{Fig:CompartmentMethod} Diagram of the method of 
node and link classifications. Open symbols ($\circ$) represent 
susceptible nodes, and solid symbols ($\bullet$) represent 
infected nodes. (a) The hub node $O$ is a susceptible node. It 
has five neighbors, in which three are infected. Therefore, this 
node is in the set $S_{k,l}$ where $k=5$ and $l=3$. (b) The link 
plotted with the thick straight line, denoted as $AB$, has an 
infected node, denoted as $A$, at its left end and a susceptible 
node, denoted as $B$, at its right end. For node $A$, in addition 
to link $AB$ it has four other links, of which two connect to 
infected nodes. While for node $B$, in addition to link $AB$ it 
has three other links, of which one connects to an infected node. 
Therefore, link $AB$ is in the set $I_{ij}S_{kl}$ with $i=4$, 
$j=2$, $k=3$, and $l=1$. Besides, node $A$ is in the set 
$I_{5,2}$ and node $B$ is in the set $S_{4,2}$.}
\end{figure}

We first classify the nodes according to their states and the 
states of their neighbors by $S_{k,l}$ and $I_{k,l}$, 
respectively, where $S_{k,l}$ ($I_{k,l}$) denotes the fraction of 
the nodes that are susceptible (infected) with $k$ neighbors 
among which $l$ are infected. Therefore, we have
\begin{equation}
\label{eq:nodes=1} \sum_{k,l}\left(S_{k,l}+I_{k,l}\right)=1.
\end{equation}
We then proceed to classify the links. The links can be easily 
classified into four types, namely SS, II, IS, and SI, where S 
and I denote the states of the two end nodes of the link, 
respectively. Further classification of the links is done by 
introducing a function $X_{ij}Y_{kl}$, where $X$ and $Y$ can be 
either S or I. Specifically, for the $X$-state node, besides its 
$Y$-state neighbor connected through the link being considered, 
it has $i$ other neighbors among which $j$ are infected; and 
similarly for the $Y$-state node. Hence, the two end nodes of a 
link belonging to $I_{ij}S_{kl}$ belong to $I_{i+1,j}$ and 
$S_{k+1,l+1}$, respectively (see 
Fig.\,\ref{Fig:CompartmentMethod}). In an undirected network, a 
link belonging to $X_{ij}Y_{kl}$ also belongs to $Y_{kl}X_{ij}$. 
Therefore, $S_{k,l}$ and $I_{k,l}$ represent different classes of 
nodes and $X_{ij}Y_{kl}$ represent different classes of links 
\cite{Notation}.

According to the conservation relations that
\begin{equation}
\label{eq:Link=1}
\sum_{i,j,k,l}\left(S_{ij}S_{kl}+I_{ij}I_{kl}+I_{ij}S_{kl}+S_{ij}I_{kl}\right)=1,
\end{equation}
and
\begin{equation}
\label{eq:AverageDegree}
\sum_{k,l}k\left(S_{k,l}+I_{k,l}\right)=\langle k \rangle,
\end{equation}
we obtain
\begin{eqnarray}
\label{eq:Relations}
\sum_{k,l}S_{ij}S_{kl}&=&(i+1-j)S_{i+1,j}\bigg{/}\sum_{k,l}{k(S_{k,l}+I_{k,l})}\nonumber\\
&=&(i+1-j)S_{i+1,j}\big/\langle k \rangle,\nonumber\\
\sum_{k,l}I_{ij}I_{kl}&=&(j+1)I_{i+1,j+1}\big/\langle k \rangle,\nonumber\\
\sum_{k,l}I_{ij}S_{kl}&=&(i+1-j)I_{i+1,j}\big/\langle k \rangle,\nonumber\\
\sum_{k,l}S_{ij}I_{kl}&=&(j+1)S_{i+1,j+1}\big/\langle k \rangle.
\end{eqnarray}
The first equation in Eq.\,(\ref{eq:Relations}) comes from the
fact that the summation of $S_{ij}S_{kl}$ with respect to all
possible $k$ and $l$ equals the fraction of links connecting an 
$S_{i+1,j}$ node at one end and an arbitrary S node at the other 
end.

Further, with the equal footing of two end nodes of a link in an
undirected network, we have
\begin{eqnarray}
\label{eq:Symmetry} S_{ij}S_{kl}=S_{kl}S_{ij},\quad
I_{ij}I_{kl}=I_{kl}I_{ij},\quad I_{ij}S_{kl}=S_{kl}I_{ij}.
\end{eqnarray}
Eqs.\,(\ref{eq:nodes=1})-(\ref{eq:Symmetry}) provide the
constraints on both node classes and link classes.

Next, we provide initial condition for the sizes of all the 
classes. Suppose initially a fraction $\varepsilon$ of nodes are 
randomly selected to be infected. Then, we have
\begin{eqnarray}
\label{eq:InitialNode} S_{k,l}&=&(1-\varepsilon)p_k {k \choose
l}\varepsilon^l(1-\varepsilon)^{k-l},\nonumber\\
I_{k,l}&=&\varepsilon p_k {k \choose
l}\varepsilon^l(1-\varepsilon)^{k-l},
\end{eqnarray}
and
\begin{align}
\label{eq:InitialLink}
S_{ij}S_{kl}&\hspace{-1mm}=\hspace{-1mm}(1-\varepsilon)^2p_{i+1,k+1}
{i \choose j}\varepsilon^j(1-\varepsilon)^{i-j}{k \choose
l}\varepsilon^l(1-\varepsilon)^{k-l},\nonumber\\
I_{ij}I_{kl}&\hspace{-1mm}=\hspace{-1mm}\varepsilon^2 p_{i+1,k+1} 
{i \choose j}\varepsilon^j(1-\varepsilon)^{i-j}{k \choose
l}\varepsilon^l(1-\varepsilon)^{k-l},\nonumber\\
I_{ij}S_{kl}&\hspace{-1mm}=\hspace{-1mm}\varepsilon(1-\varepsilon)p_{i+1,k+1}
{i \choose j}\varepsilon^j(1-\varepsilon)^{i-j}{k \choose
l}\varepsilon^l(1-\varepsilon)^{k-l},\nonumber\\
S_{ij}I_{kl}&\hspace{-1mm}=\hspace{-1mm}(1-\varepsilon)\varepsilon
p_{i+1,k+1} {i \choose j}\varepsilon^j(1-\varepsilon)^{i-j}{k
\choose
l}\varepsilon^l(1-\varepsilon)^{k-l}.\nonumber\\\hspace{-4mm}
\end{align}
We take $S_{ij}S_{kl}$ as an example to show the reasoning of the
above equations. An $S_{ij}S_{kl}$ link has two end nodes with 
degree $i+1$ and $k+1$, respectively. The probability of picking 
such a link therefore equals $p_{i+1,k+1}$, and the probability 
that both of its end nodes being susceptible equals 
$(1-\varepsilon)^2$. Furthermore, the probability that among the 
$i$ [$k$] links $j$ [$l$] of them connect to infected nodes 
equals ${i \choose j}\varepsilon^j(1-\varepsilon)^{(i-j)}$ [${k 
\choose l}\varepsilon^l(1-\varepsilon)^{(k-l)}$].

After obtaining the constraints and the initial conditions for 
all the classes, we now provide the corresponding ordinary 
differential equation (ODE) to describe the time evolution of the 
system. To simplify the discussion, we separate the whole process 
into three sub-processes, named ``Recovery Process", ``Infection 
Process" and ``Rewiring Process", respectively. Since during a 
very short time the mutual influences of these three 
sub-processes could be ignored, for each class we shall firstly 
calculate its variations in the three sub-processes, 
respectively, and then sum them up to obtain the variation of the 
whole process. Denoting the differential operators of the ODEs 
describing the three sub-processes and the whole process as 
``$\textrm{d}^\textrm{R}/\textrm{d}t$", 
``$\textrm{d}^\textrm{I}/\textrm{d}t$", 
``$\textrm{d}^\textrm{W}/\textrm{d}t$", and 
``$\textrm{d}/\textrm{d}t$" respectively, we have 
$\textrm{d}/\textrm{d}t=\textrm{d}^\textrm{R}/\textrm{d}t+\textrm{d}^\textrm{I}/\textrm{d}t+\textrm{d}^\textrm{W}/\textrm{d}t$.

Firstly, for the Recovery Process, there are two different 
situations that may cause the variations in the size of the node 
classes and link classes, respectively. For the node classes 
$S_{k,l}$ and $I_{k,l}$, the first situation is the change of the 
infection state that happens on the nodes belonging to $S_{k,l}$ 
and $I_{k,l}$ (for example, the node $O$ in Fig. 
\ref{Fig:CompartmentMethod}(a)), and the second one is the change 
of the infection state that happens on the neighboring nodes of 
those nodes belonging to $S_{k,l}$ and $I_{k,l}$ (for example, 
the neighbors of node $O$ in Fig. 
\ref{Fig:CompartmentMethod}(a)). Similarly, for the link class 
$X_{ij}Y_{kl}$, where $X$ and $Y$ can be either $S$ or $I$, the 
first situation is the change of the infection state that happens 
on the nodes at the ends of the links belonging to $X_{ij}Y_{kl}$ 
(for example, the nodes $A$ and $B$ in Fig. 
\ref{Fig:CompartmentMethod}(b)), and the second one is the change 
of the infection state that happens on the neighboring nodes of 
those at the ends of the links belonging to $X_{ij}Y_{kl}$ (for 
example, the nodes other than $A$ and $B$ in Fig. 
\ref{Fig:CompartmentMethod}(b)). To distinct the two situations, 
in the formalism describing the Recovery Process provided below, 
the terms corresponding to the first situation are labelled with 
``\{Self\}" on the right side of the equations: 
\begin{flalign}
\label{eq:Recovery-Skl}
\frac{\textrm{d}^\textrm{R}S_{k,l}}{\textrm{d}t}=&\gamma I_{k,l}&
\textrm{\{Self\}}\nonumber\\
&+\gamma[(l+1)S_{k,l+1}-lS_{k,l}],
\end{flalign}
\begin{flalign}
 \label{eq:Recovery-Ikl} 
\frac{\textrm{d}^\textrm{R}I_{k,l}}{\textrm{d}t}=&-\gamma I_{k,l}&
\textrm{\{Self\}}\nonumber\\
&+\gamma[(l+1)I_{k,l+1}-lI_{k,l}],
\end{flalign}
and
\begin{flalign}
\label{eq:Recovery-Link}
\frac{\textrm{d}^\textrm{R}S_{ij}S_{kl}}{\textrm{d}t}=&\gamma
I_{ij}S_{kl}+\gamma S_{ij}I_{kl}&
\textrm{\{Self\}}\nonumber\\
&-\gamma jS_{ij}S_{kl}+\gamma(j+1)S_{i(j+1)}S_{kl}\nonumber\\
&-\gamma lS_{ij}S_{kl}+\gamma(l+1)S_{ij}S_{k(l+1)},
\end{flalign}
\begin{flalign}
\frac{\textrm{d}^\textrm{R}I_{ij}I_{kl}}{\textrm{d}t}=&-\gamma 
I_{ij}I_{kl}-\gamma I_{ij}I_{kl}&
\textrm{\{Self\}}\nonumber\\
&-\gamma jI_{ij}I_{kl}+\gamma(j+1)I_{i(j+1)}I_{kl}\nonumber\\
&-\gamma lI_{ij}I_{kl}+\gamma(l+1)I_{ij}I_{k(l+1)},\end{flalign}
\begin{flalign}
\frac{\textrm{d}^\textrm{R}I_{ij}S_{kl}}{\textrm{d}t}=&-\gamma 
I_{ij}S_{kl}+\gamma I_{ij}I_{kl}&
\textrm{\{Self\}}\nonumber\\
&-\gamma jI_{ij}S_{kl}+\gamma(j+1)I_{i(j+1)}S_{kl}\nonumber\\
&-\gamma lI_{ij}S_{kl}+\gamma(l+1)I_{ij}S_{k(l+1)},\end{flalign}
\begin{flalign}
\frac{\textrm{d}^\textrm{R}S_{ij}I_{kl}}{\textrm{d}t}=&-\gamma 
S_{ij}I_{kl}+\gamma I_{ij}I_{kl}&
\textrm{\{Self\}}\nonumber\\
&-\gamma jS_{ij}I_{kl}+\gamma(j+1)S_{i(j+1)}I_{kl}\nonumber\\
&-\gamma lS_{ij}I_{kl}+\gamma(l+1)S_{ij}I_{k(l+1)}.
\end{flalign}
Diagrams of the Recovery Process described by the above equations 
are depicted in Figs.\,\ref{Fig:Diagram}\,(a)-(c).

Similar with the Recovery Process, for the Infection Process 
there are also two different situations that may cause the 
variations of the sizes of the node classes $S_{k,l}$ and 
$I_{k,l}$ and of the link class $X_{ij}Y_{kl}$. The first 
situation is the change of the infection state that happens on 
the nodes belonging to $S_{k,l}$ and $I_{k,l}$ or at the ends of 
$X_{ij}Y_{kl}$, and the second one is the change of the infection 
state that happens on the neighbors of the nodes referred in the 
first situation. Now, we give the formalism for the Infection 
Process. Again, we label ``\{self\}" on the right side of the 
terms corresponding to the first situation. To facilitate the 
understanding of the terms describing the second situation, for 
each term we list the corresponding process it describes in 
braces on its right side, where ``$\sim$" denotes the terms under 
derivation on the left side of the corresponding equations. 
Moreover, we provide a detailed explanation of the second term on 
the right side of Eq.\,(\ref{eq:Infection-Skl}) in Appendix 
\ref{App:Deviation-dIdtSkl}.
\begin{flalign}
\label{eq:Infection-Skl}
\frac{\textrm{d}^\textrm{I}S_{k,l}}{\textrm{d}t}=&-\beta
lS_{k,l}&
\textrm{\{Self\}}\nonumber\\
&-\beta\sum_{i',j'}{j'S_{i'j'}S_{(k-1)l}}\langle k \rangle\nonumber\\
&+\beta\sum_{i',j'}{j'S_{i'j'}S_{(k-1)(l-1)}}\langle k 
\rangle,\end{flalign}
\begin{flalign}
\label{eq:Infection-Ikl} 
\frac{\textrm{d}^\textrm{I}I_{k,l}}{\textrm{d}t}=&\beta lS_{k,l}&
\textrm{\{Self\}}\nonumber\\
&-\beta\sum_{i',j'}{(j'+1)S_{i'j'}I_{(k-1)l}}\langle k \rangle\nonumber\\
&+\beta\sum_{i',j'}{(j'+1)S_{i'j'}S_{(k-1)(l-1)}}\langle k
\rangle,
\end{flalign}
and
\begin{flalign}
\label{eq:Infection-SijSkl}
&\frac{\textrm{d}^\textrm{I}S_{ij}S_{kl}}{\textrm{d}t}=-\beta
jS_{ij}S_{kl}-\beta lS_{ij}S_{kl}&
\textrm{\{Self\}}\nonumber\\
&-\beta (i-j)\frac{\sum\limits_{k',l'}S_{ij}S_{k'l'}l'\langle k
\rangle}{(i-j+1)S_{i+1,j}}S_{ij}S_{kl}&\{{\scriptscriptstyle
{\displaystyle\sim}\rightarrow S_{i(j+1)}S_{kl} }\}\nonumber\\
&-\beta 
(i-j+1)\frac{\sum\limits_{k',l'}S_{i(j-1)}S_{k'l'}l'\langle k 
\rangle}{(i-j+2)S_{i+1,j-1}}S_{i(j-1)}S_{kl}\hspace{-8mm}\nonumber\\
&&\{{\scriptscriptstyle
S_{i(j-1)}S_{kl}\rightarrow {\displaystyle\sim} }\}\nonumber\\
&-\beta (k-l)\frac{\sum\limits_{i',j'}j'S_{i'j'}S_{kl}\langle k 
\rangle}{(k-l+1)S_{k+1,l}}S_{ij}S_{kl}&\{{\scriptscriptstyle
{\displaystyle\sim}\rightarrow S_{ij}S_{k(l+1)} }\}\nonumber\\
&-\beta
(k-l+1)\frac{\sum\limits_{i',j'}j'S_{i'j'}S_{k(l-1)}\langle k
\rangle}{(k-l+2)S_{k+1,l-1}}S_{ij}S_{k(l-1)},\hspace{-8mm}\nonumber\\
&&\{{\scriptscriptstyle S_{ij}S_{k(l-1)}\rightarrow 
{\displaystyle\sim} }\}
\end{flalign}
\begin{flalign}
\label{eq:Infection-IijIkl} 
&\frac{\textrm{d}^\textrm{I}I_{ij}I_{kl}}{\textrm{d}t}=\beta 
(j+1)S_{ij}I_{kl}+\beta (l+1)I_{ij}S_{kl}\hspace{-4mm}&
\textrm{\{Self\}}\nonumber\\
&-\beta \frac{\sum\limits_{k',l'}I_{i(j+1)}S_{k'l'}(l'+1)\langle 
k 
\rangle}{I_{i+1,j+1}}I_{ij}I_{kl}\hspace{-4mm}&\{{\scriptscriptstyle
{\displaystyle\sim}\rightarrow I_{i(j+1)}I_{kl} }\}\nonumber\\
&+\beta \frac{\sum\limits_{k',l'}I_{ij}S_{k'l'}(l'+1)\langle k 
\rangle}{I_{i+1,j}}I_{i(j-1)}I_{kl}\hspace{-4mm}&\{{\scriptscriptstyle
I_{i(j-1)}I_{kl}\rightarrow {\displaystyle\sim} }\}\nonumber\\
&-\beta \frac{\sum\limits_{i',j'}(j'+1)S_{i'j'}I_{k(l+1)}\langle 
k 
\rangle}{I_{k+1,l+1}}I_{ij}I_{kl}\hspace{-4mm}&\{{\scriptscriptstyle
{\displaystyle\sim}\rightarrow I_{ij}I_{k(l+1)} }\}\nonumber\\
&+\beta \frac{\sum\limits_{i',j'}(j'+1)S_{i'j'}I_{kl}\langle k 
\rangle}{I_{k+1,l}}I_{ij}I_{k(l-1)},\hspace{-4mm}&\{{\scriptscriptstyle
I_{ij}I_{k(l-1)}\rightarrow {\displaystyle\sim} }\}\nonumber\\
\end{flalign}
\begin{flalign}
\label{eq:Infection-IijSkl}
&\frac{\textrm{d}^\textrm{I}I_{ij}S_{kl}}{\textrm{d}t}=\beta
jS_{ij}S_{kl}-\beta (l+1)I_{ij}S_{kl}&
\textrm{\{Self\}}\nonumber\\
&-\beta (i-j)\frac{\sum\limits_{k',l'}I_{ij}S_{k'l'}(l'+1)\langle
k \rangle}{(i-j+1)I_{i+1,j}}I_{ij}S_{kl}&\{{\scriptscriptstyle
{\displaystyle\sim}\rightarrow I_{i(j+1)}S_{kl} }\}\nonumber\\
&+\beta
(i-j+1)\frac{\sum\limits_{k',l'}I_{i(j-1)}S_{k'l'}(l'+1)\langle k
\rangle}{(i-j+2)I_{i+1,j-1}}I_{i(j-1)}S_{kl}\hspace{-16mm}\nonumber\\
&&\{{\scriptscriptstyle
I_{i(j-1)}S_{kl}\rightarrow {\displaystyle\sim} }\}\nonumber\\
&-\beta \frac{\sum\limits_{i',j'}j'S_{i'j'}S_{k(l+1)}\langle k 
\rangle}{S_{k+1,l+1}}I_{ij}S_{kl}&\{{\scriptscriptstyle
{\displaystyle\sim}\rightarrow I_{ij}S_{k(l+1)} }\}\nonumber\\
&+\beta \frac{\sum\limits_{i',j'}j'S_{i'j'}S_{kl}\langle k
\rangle}{S_{k+1,l}}I_{ij}S_{k(l-1)},&\{{\scriptscriptstyle
I_{ij}S_{k(l-1)}\rightarrow{\displaystyle\sim} }\}\nonumber\\
\end{flalign}
\begin{flalign}
\label{eq:Infection-SijIkl} 
&\frac{\textrm{d}^\textrm{I}S_{ij}I_{kl}}{\textrm{d}t}=-\beta 
(j+1)S_{ij}I_{kl}+\beta lS_{ij}S_{kl}&\textrm{\{Self\}}\nonumber\\
&-\beta \frac{\sum\limits_{k',l'}S_{i(j+1)}S_{k'l'}l'\langle k 
\rangle}{S_{i+1,j+1}}S_{ij}I_{kl}&\{{\scriptscriptstyle
{\displaystyle\sim}\rightarrow S_{i(j+1)}I_{kl} }\}\nonumber\\
&+\beta \frac{\sum\limits_{k',l'}S_{ij}S_{k'l'}l'\langle k 
\rangle}{S_{i+1,j}}S_{i(j-1)}I_{kl}&\{{\scriptscriptstyle
S_{i(j-1)}I_{kl}\rightarrow {\displaystyle\sim} }\}\nonumber\\
&-\beta (k-l)\frac{\sum\limits_{i',j'}(j'+1)S_{i'j'}I_{kl}\langle 
k \rangle}{(k-l+1)I_{k+1,l}}S_{ij}I_{kl}&\{{\scriptscriptstyle
{\displaystyle\sim}\rightarrow S_{ij}I_{k(l+1)} }\}\nonumber\\
&+\beta
(k-l+1)\frac{\sum\limits_{i',j'}(j'+1)S_{i'j'}I_{k(l-1)}\langle k
\rangle}{(k-l+2)I_{k+1,l-1}}S_{ij}I_{k(l-1)}.\hspace{-16mm}\nonumber\\
&&\{{\scriptscriptstyle
S_{ij}I_{k(l-1)}\rightarrow {\displaystyle\sim}}\}\nonumber\\
\end{flalign}

Diagrams of the Infection process described by the above 
equations are depicted in Figs.\,\ref{Fig:Diagram}\,(d)-(f).

Finally, for the Rewiring process, it is composed of two 
sub-processes. One is the link breaking process, where some 
susceptible nodes break their links connecting to infected 
neighbors; and the other is the link attachment process, where 
susceptible nodes rewire their broken links to other susceptible 
nodes so as to form new connections. In the following, we give 
the formalism for the Rewiring Process. We label ``\{break\}" on 
the right side of all the terms describing the link breaking 
process; for each term describing the link attachment process we 
show the corresponding process it describes in braces on its 
right side with ``$\sim$" having the same meaning as that in the 
Infection Process.
\begin{flalign}
\label{eq:Rewiring-Skl}
\frac{\textrm{d}^\textrm{W}S_{k,l}}{\textrm{d}t}=&-\omega
lS_{k,l}+\omega(l+1)S_{k,l+1}&
\textrm{\{Break\}}\nonumber\\
&-\omega(IS)\frac{S_{k,l}}{S}+\omega(IS)\frac{S_{k-1,l}}{S},\end{flalign}
\begin{flalign}
\label{eq:Rewiring-Ikl} 
\frac{\textrm{d}^\textrm{W}I_{k,l}}{\textrm{d}t}=&-\omega 
(k-l)I_{k,l}+\omega(k-l+1)I_{k+1,l},&
\textrm{\{Break\}}\nonumber\\
\end{flalign}
\begin{flalign}
\label{eq:Rewiring-SijSkl}
\frac{\textrm{d}^\textrm{W}S_{ij}S_{kl}}{\textrm{d}t}=&-\omega
jS_{ij}S_{kl}+\omega (j+1)S_{i(j+1)}S_{kl}&
\textrm{\{Break\}}\nonumber\\
&-\omega lS_{ij}S_{kl}+\omega (l+1)S_{ij}S_{k(l+1)}&\textrm{\{Break\}}\nonumber\\
&-\omega|IS|\frac{S_{ij}S_{kl}\langle k
\rangle}{|S|}&\hspace{-6mm}\{{\scriptscriptstyle
{\displaystyle\sim}
\rightarrow S_{(i+1)j}I_{kl}}\}\nonumber\\
&+\omega|IS|\frac{S_{(i-1)j}S_{kl}\langle k
\rangle}{|S|}&\hspace{-6mm}\{{\scriptscriptstyle S_{(i-1)j}I_{kl}
\rightarrow {\displaystyle\sim}}\}\nonumber\\
&-\omega|IS|\frac{S_{ij}S_{kl}\langle k
\rangle}{|S|}&\hspace{-6mm}\{{\scriptscriptstyle
{\displaystyle\sim}
\rightarrow S_{ij}I_{(k+1)l}}\}\nonumber\\
&+\omega|IS|\frac{S_{ij}S_{(k-1)l}\langle k
\rangle}{|S|}&\hspace{-6mm}\{{\scriptscriptstyle S_{ij}I_{(k-1)l}
\rightarrow {\displaystyle\sim}}\}\nonumber\\
&+\omega|IS_{kl}|\frac{S_{i,j}}{|S|}+\omega|S_{ij}I|\frac{S_{k,l}}{|S|},\nonumber\\
\end{flalign}
\begin{flalign}
\label{eq:Rewiring-IijIkl}
\frac{\textrm{d}^\textrm{W}I_{ij}I_{kl}}{\textrm{d}t}=&-\omega(i-j)I_{ij}I_{kl}&\textrm{\{Break\}}\nonumber\\
&+\omega(i-j+1)I_{(i+1)j}I_{kl}&\textrm{\{Break\}}\nonumber\\
&-\omega(k-l)I_{ij}I_{kl}&\textrm{\{Break\}}\nonumber\\
&+\omega(k-l+1)I_{ij}I_{(k+1)l},&\textrm{\{Break\}}\nonumber\\
\end{flalign}
\begin{flalign}
\label{eq:Rewiring-IijSkl}
\frac{\textrm{d}^\textrm{W}I_{ij}S_{kl}}{\textrm{d}t}=&-\omega
I_{ij}S_{kl}-\omega(i-j)I_{ij}S_{kl}&\textrm{\{Break\}}\nonumber\\
&+\omega(i-j+1)I_{(i+1)j}S_{kl}&\textrm{\{Break\}}\nonumber\\
&-\omega lI_{ij}S_{kl}+\omega(l+1)I_{ij}S_{k(l+1)}&\textrm{\{Break\}}\nonumber\\
&-\omega|IS|\frac{I_{ij}S_{kl}\langle k
\rangle}{|S|}&\hspace{-6mm}\{{\scriptscriptstyle
{\displaystyle\sim}
\rightarrow I_{ij}S_{(k+1)l}}\}\nonumber\\
&+\omega|IS|\frac{I_{ij}S_{(k-1)l}\langle k
\rangle}{|S|},&\hspace{-6mm}\{{\scriptscriptstyle I_{ij}S_{(k-1)l}
\rightarrow {\displaystyle\sim}}\}\nonumber\\
\end{flalign}
\begin{flalign}
\label{eq:Rewiring-SijIkl}
\frac{\textrm{d}^\textrm{W}S_{ij}I_{kl}}{\textrm{d}t}=&-\omega
S_{ij}I_{kl}-\omega jS_{ij}I_{kl}&\textrm{\{Break\}}\nonumber\\
&+\omega(j+1)S_{i(j+1)}I_{kl}&\textrm{\{Break\}}\nonumber\\
&-\omega (k-l)S_{ij}I_{kl}&\textrm{\{Break\}}\nonumber\\
&+\omega(k-l+1)S_{ij}I_{k(l+1)}&\textrm{\{Break\}}\nonumber\\
&-\omega|IS|\frac{S_{ij}I_{kl}\langle k
\rangle}{|S|}&\hspace{-6mm}\{{\scriptscriptstyle
{\displaystyle\sim}
\rightarrow S_{(i+1)j}I_{kl}}\}\nonumber\\
&+\omega|IS|\frac{S_{(i-1)j}I_{kl}\langle k
\rangle}{|S|},&\hspace{-6mm}\{{\scriptscriptstyle S_{(i-1)j}I_{kl}
\rightarrow {\displaystyle\sim}}\}\nonumber\\
\end{flalign}where $|IS|=\sum_{i,j,k,l}{I_{ij}S_{kl}}$,
$|IS_{kl}|=\sum_{i,j}{I_{ij}S_{kl}}$, 
$|S_{ij}I|=\sum_{i,j}S_{ij}I_{kl}$ and $|S|=\sum_{k,l}S_{k,l}$.

\begin{figure*}[t]
\includegraphics[width=0.85\linewidth]{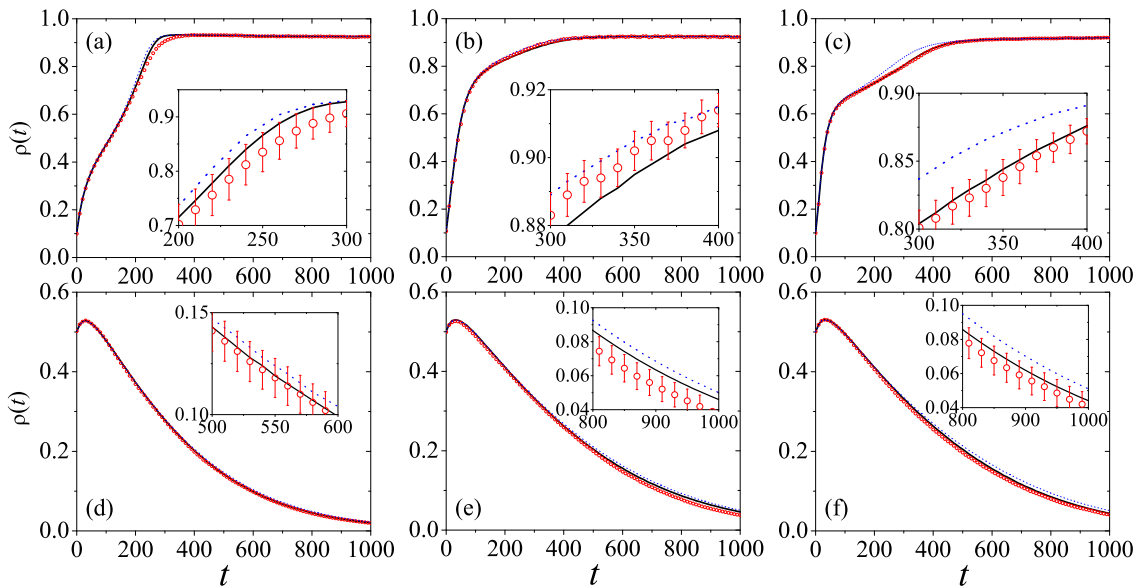}
\caption{\label{Fig:Uncorrelated} (Color online) Time evolution 
of the infected fraction $\rho(t)$ for the cases of 
$\gamma=0.005$, $\omega=0.02$, (a)-(c) $\beta=0.06$, 
$\varepsilon=0.1$, where the systems eventually evolve to the 
endemic state, and (d)-(f) $\beta=0.01$, $\varepsilon=0.5$ where 
the systems eventually evolve to the disease-free state. The 
black solid curves are obtained from our link-based method; the 
blue dotted curves are obtained from the node-based method in 
Ref.~\cite{Marceau:2010}; and the red circle symbols are obtained 
from simulations. Left, middle, and right panels correspond 
to different cases for $k$-regular network, Poisson network, and 
power-law network, respectively. The average degrees of all the 
networks are the same, $\langle k \rangle=2$. For the power-Law 
network, $k_T=13$ and $\tau=2.019$.}
\end{figure*}

The terms describe the link breaking process is relatively easy 
to understand. For example, the first term on the right side of 
Eq.\,(\ref{eq:Rewiring-IijIkl}) means: For the $I_{ij}I_{kl}$ 
link, the $I_{(i+1),(j+1)}$ node at one of its ends has $(i-j)$ 
neighboring susceptible nodes. Thus, the probability that the 
node looses a neighbor in a time step equals $\omega(i-j)$. Once 
this event happens, an $I_{ij}I_{kl}$ link changes to an 
$I_{(i-1)j}I_{kl}$ link. Hence, for all the $I_{ij}I_{kl}$ links 
the probability that one of them changes to the 
$I_{(i-1)j}I_{kl}$ link in one time step equals 
$\omega(i-j)I_{ij}I_{kl}$. The link attachment process is a bit 
more complicated. Specifically, the last term in 
Eq.\,(\ref{eq:Rewiring-SijSkl}) denotes the probability of the 
event that a link previously having the $S_{i+1,j+1}$ node at one 
end and an infected node at the other end now becomes the 
$S_{ij}S_{kl}$ links by rewiring itself from the infected node to 
the $S_{k,l}$ node. To help a better understanding, we give a 
detailed explanation of the fifth term on the right side of 
Eq.\,(\ref{eq:Rewiring-SijSkl}) in the Appendix 
\ref{App:Deviation-dWdtSijSkl}. Diagrams of the Rewiring process 
described by the above equations are depicted in 
Figs.\,\ref{Fig:Diagram}\,(g)-(k).

Note that Eqs.\,(\ref{eq:Recovery-Skl})-(\ref{eq:Recovery-Ikl}) 
and (\ref{eq:Rewiring-Skl})-(\ref{eq:Rewiring-Ikl}) do not have 
any link term $X_{ij}Y_{kl}$, which means in the Recovery Process 
and Rewiring Process, the link classification information is 
\emph{not} directly utilized to calculate the variations of the 
sizes of the node classes. However, 
Eqs.\,(\ref{eq:Infection-Skl})-(\ref{eq:Infection-Ikl}) do 
contain the information of link classes. The information of link 
classification is utilized to calculate the variations of the 
sizes of the node classes in the Infection Process. Since the 
link classification method provides more detailed information of 
the network structure than the node classification, the accuracy 
in predicting the time evolution of the node states could be 
improved by employing our method, as shown in the following 
section.

\section{Simulation results}
\label{sec:Results}
\subsection{Uncorrelated networks}

In the simulations, unless otherwise specified, all the results 
are obtained based on networks with size $N=10000$ and averaged 
on $1000$ different realizations.

In this subsection, we will compare our link-based method to the 
node-based method  proposed in Ref.~\cite{Marceau:2010}. Specifically, we adopt 
three kinds of initial degree distribution: (i) $k$-regular, 
where the degrees of all the nodes are the same as $k$. (ii) 
Poisson distribution, where $p_k=z^ke^{-z}/k!$ with $z=\langle k 
\rangle$ being the average degree. This distribution can be 
achieved from a random-graph model in the limit 
$N\rightarrow\infty$. (iii) Truncated power-law distribution, 
where $p_k=ck^{-\tau}$ when $0<k<k_T$ and $p_k=0$ otherwise. For 
convenience of discussion, we refer to the three kinds of 
networks as $k$-regular network, Poisson network, and power-law 
network, respectively.

We start from the case that the initial network topology has no 
degree correlation. Figure\,\ref{Fig:Uncorrelated} shows the 
infected fraction $\rho(t)$ for the three kinds of networks, 
respectively. We observe that both our link-based formalism 
(solid black curve) and the existing node-based formalism 
\cite{Marceau:2010} (blue dotted curve) have high accuracy in 
predicting $\rho(t)$. However, detailed plots in the insets of 
Fig.\,\ref{Fig:Uncorrelated} show that link-based formalism 
offers better approximation, except for panel (b) where their 
performances are similar. As mentioned earlier, such improvements 
are a result of utilizing the link information, which contains 
more details about the network structure compared with the pure 
node-based method.

When the proposed formalism is applied on networks with nonempty 
degree classes, the total number of differential equations is 
$3(k_\mathrm{max}+1)(k_\mathrm{max}+2)+3(k_\mathrm{max}+1)^2(k_\mathrm{max}+2)^2$ 
with $k_\mathrm{max}$ being the largest degree, which grows with 
the fourth power of the largest nodal degree. For the node-based 
method used in \cite{Marceau:2010}, the number of the 
differential equations is $(k_\mathrm{max}+1)(k_\mathrm{max}+2)$, 
which is more efficient. Therefore, on uncorrelated networks 
where the interest is mainly on the infected fraction rather than 
network topology, the node based method can be a good 
approximation to describe the system. For the case of correlated 
networks, however, the following subsection will show that our 
method can obtain prediction results with high accuracy.

\subsection{Correlated Networks}

Now we study how the degree correlation evolves in an adaptive
network. We focus on the case where the system in the endemic 
state could evolve sufficiently and the evolution of the degree 
correlation is evident. For convenience in presenting the degree 
correlation, we introduce a relative correlation function defined 
by $f_{i,k}=p_{i,k}/q_iq_k$. Obviously, when $f_{i,k}=1$, for all 
the $i$ and $k$ the network is uncorrelated. For positive 
(negative) correlated networks, a smaller difference between the 
values of $i$ and $k$ leads to a larger (smaller) value of 
$f_{i,k}$.

In our formalisms, the initial joint degree distribution function 
$p_{i,k}$ can be arbitrary. To verify the formalism, we need to 
construct networks with a given degree distribution of which the 
initial degree correlations are tunable so as to satisfy the 
given $p_{i,k}$. A method to construct networks with arbitrarily 
given degree correlations is provided in Ref.\,\cite{Weber:2007}, 
which also proposes an approach for generating a relative 
correlation function $f_{i,k}$ satisfying a pre-assigned score of 
factor $r$. Specifically, suppose that the function $f_{i,k}$ is 
required to satisfy an average nearest-neighbor function 
$k_{nn}(k)$ with $k_{nn}(k) \propto k^\alpha$, where $k_{nn}(k)$ 
denotes the mean value of the average degree of the nearest 
neighbors of all the $k$-degree nodes. Then, when $\alpha>0$ the 
network possesses a positive degree correlation and a larger 
$\alpha$ corresponds to a larger score of $r$, and vice versa. It 
is shown that when $f_{i,k}$ is obtained from the following 
equation, the condition of $k_{nn}(k) \propto k^\alpha$ can be 
satisfied \cite{Weber:2007}:
\begin{equation}
\label{eq:fik} f_{i,k}=1+\frac{1}{\langle k^{\alpha}
\rangle_\textrm{e}}\frac{(i^\alpha-\langle k^{\alpha+1}
\rangle_\textrm{e})(k^\alpha-\langle k^{\alpha+1}
\rangle_\textrm{e})}{\langle k^{\alpha+1}
\rangle_\textrm{e}/\langle k \rangle_\textrm{e}-\langle 
k^{\alpha} \rangle_\textrm{e}},
\end{equation}
where $\langle\bullet\rangle_\textrm{e}=\sum_k{(\bullet\times
q_k)}$. Therefore, by tuning $\alpha$ we may get the desired
function $f_{i,k}$ from which a pre-assigned function $k_{nn}(k)$
and correspondingly the desired $r$ score can be obtained. 

\begin{figure}
\includegraphics[width=1.0\linewidth]{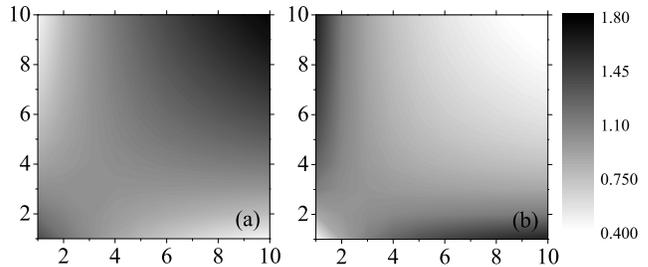}
\caption{\label{Fig:fij} Gray-scale plots of $f_{i,k}$ obtained 
from Eq. (\ref{eq:fik}), which satisfy the conditions of (a) 
$r=0.3$ and (b) $r=-0.3$.}
\end{figure}

\begin{figure*}
\includegraphics[width=1.0\linewidth]{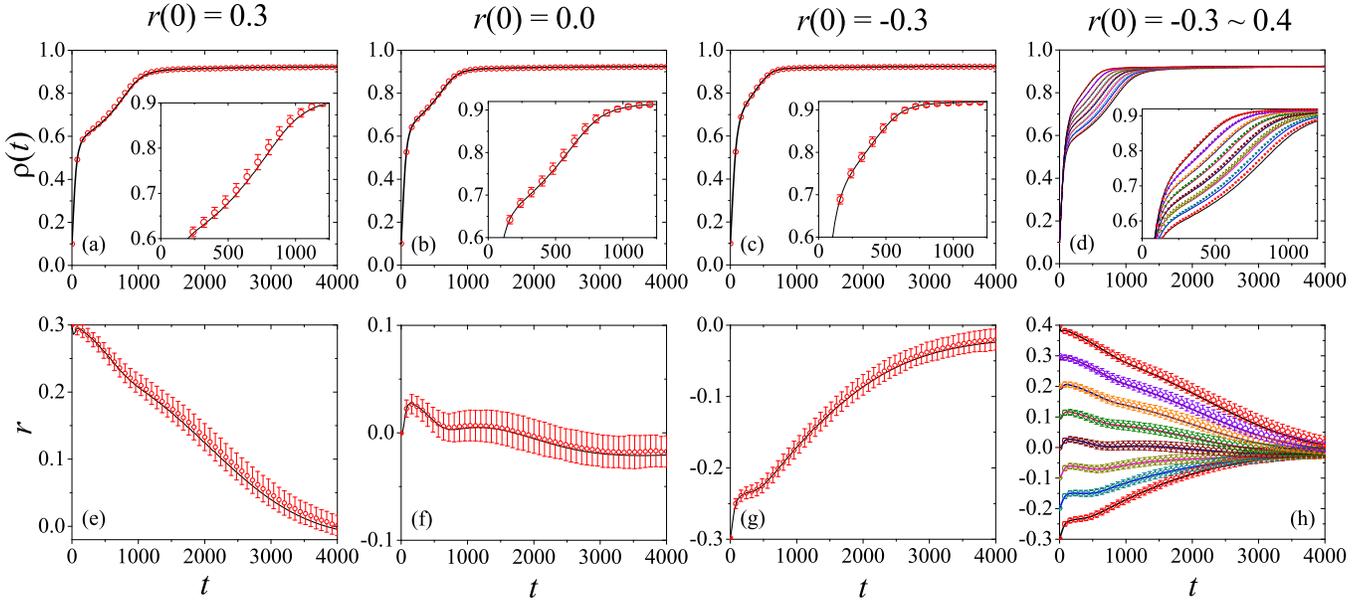}
\caption{\label{Fig:Correlated-ir} (Color online) The evolution 
of i) $\rho(t)$ for (a) $r(0)=0.3$, (b) $r(0)=0$, (c) 
$r(0)=-0.3$; and ii) $r(t)$ for (e) $r(0)=0.3$, (f) $r(0)=0$, (g) 
$r(0)=-0.3$ respectively, where the initial power-law degree 
distribution is the same as that in Fig.\,\ref{Fig:Uncorrelated}. 
Panels (d) and (h) show the evolution of $\rho(t)$ and $r(t)$ for 
different initial $r(0)$. Specifically, in panel (d), curves from 
above to bottom correspond to the cases of $r(0)=-0.3, \cdots, 
0.4$, respectively. Symbols represent the simulation results and 
the curves are obtained from the theoretical formalism. Moreover, 
the insets of panels (a)\,-\,(d) show the zoom in results for the 
middle stage of the evolution of the corresponding panels. For 
clarity, the error bars are only shown in the insets of panels 
(a)\,-\,(c), while the sizes of other ones are conform with them. 
The parameters of the epidemic model are $\gamma=0.0025$, 
$\beta=0.03$ and $\omega=0.01$.}
\end{figure*}

\begin{figure*}
\includegraphics[width=0.8\linewidth]{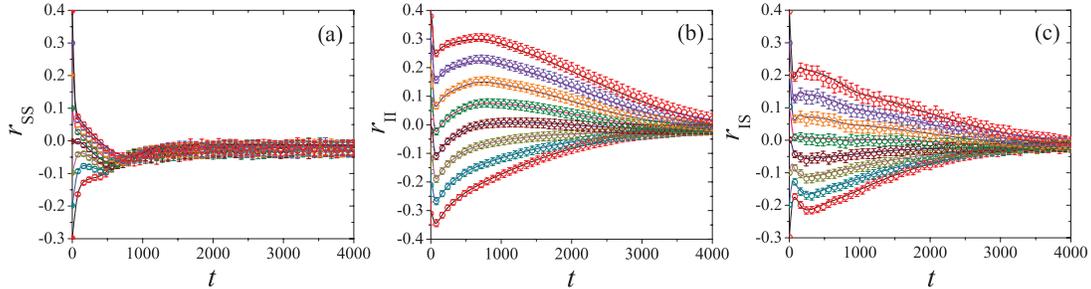}
\caption{\label{Fig:Correlated-rSSrIIrIS} (Color online) The 
evolution of the correlation measurements of (a) 
$r_\textrm{SS}(t)$, (b) $r_\textrm{II}(t)$, and (c) 
$r_\textrm{IS}(t)$ for different initial degree correlations. In 
all the panels, curves from above to below correspond to the cases 
of $r(0)=0.4, 0.3, 0.2, 0.1, 0, -0.1, -0.2, -0.3$, respectively. 
Other parameters are the same as those in 
Fig.\,\ref{Fig:Correlated-ir}.}
\end{figure*}

As an example, we show the values of $f_{i,k}$ on the 
degree-degree plane that are generated by the method introduced 
above for the cases of $r=0.3$ and $-0.3$ in 
Figs.\,\ref{Fig:fij}\,(a) and (b), respectively. For the case of 
$r=0.3$, $f_{i,k}$ with values larger than $1$ are mostly close 
to the diagonal line especially for the two ends, while for the 
case of $r=-0.3$, high values of $f_{i,k}$ dominate in the far 
off-diagonal region.

Since the degree correlation of a degree-regular network is 
trivial and the degrees of Possion distribution are distributed 
in a narrow range that largely restricts the range of $r$, we 
focus our attention on power-law networks. Figures 
\ref{Fig:Correlated-ir}\,(a)-(d) show the time evolution 
$\rho(t)$ for different values of $r(0)$ ranging from $-0.3$ to 
$0.4$, where all the cases are eventually evolve to the endemic 
state. Detailed plots in Figs.\,\ref{Fig:Correlated-ir}\,(a)-(c) 
show that the results obtained from the theoretical formalism 
(curves) have a good match with the simulation results (symbols). 
Moreover, as shown in Fig.\,\ref{Fig:Correlated-ir}\,(d), in the 
middle stage of evolution, cases with smaller $r(0)$ always have 
higher $\rho(t)$ than those with larger $r(0)$, meaning that 
negative initial correlation favors the spreading of infection in 
this case. Figures \ref{Fig:Correlated-ir}\,(e)-(h) show the 
evolution of $r(t)$ for the corresponding cases in (a)-(d), 
respectively. Figure \ref{Fig:Correlated-ir}\,(h) shows that 
$r(t)$ moves gradually towards zero in all the cases, which 
indicates that the memory of the initial degree correlation fades 
out since the nodal degree is not discriminated during the 
Rewiring Process. Finally, for all the cases, $\rho(t)$ converge 
to the same value.

\begin{figure}
\includegraphics[width=1.0\linewidth]{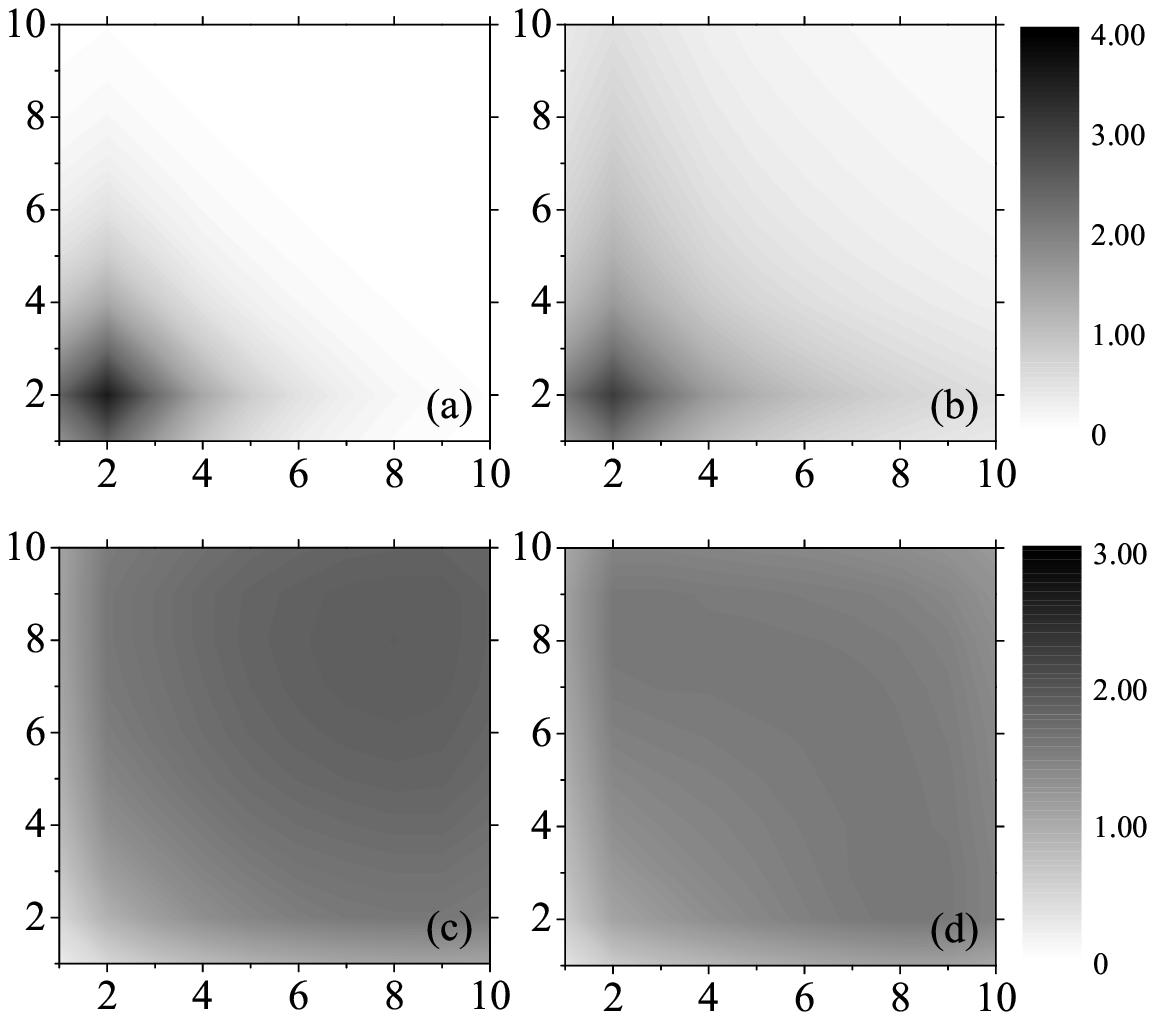}
\caption{\label{Fig:Correlated-GrayPlot} Gray-scale plots the 
ratio $p_{i,k}^\textrm{SS}(t=60)/p_{i,k}^\textrm{SS}(t=0)$ for 
the cases of (a) $r(0)=0.3$ and (b) $r(0)=-0.3$, and the ratio 
$p_{i,k}^\textrm{II}(t=60)/p_{i,k}^\textrm{II}(t=0)$ for the 
cases of (c) $r(0)=0.3$ and (d) $r(0)=-0.3$ on the degree-degree 
plane. Other parameters are the same as those in 
Fig.\,\ref{Fig:Correlated-ir}.}
\end{figure}

To study the degree correlation more carefully, we further
introduce several measures according to the nodes states at the
ends of edges. Specifically, we use $r_\textrm{SS}$, 
$r_\textrm{II}$, $r_\textrm{IS}$ to measure the degree 
correlations for the links whose ends are attached by two S 
nodes, two I nodes, and one S node and one I node, respectively. 
By using the Pearson correlation, these measures are defined as
\begin{equation}
r_\textrm{XY}=\frac{1}{\sigma_e^\textrm{X}\sigma_e^\textrm{Y}}\sum_{i,k}(p_{i,k}^\textrm{XY}-q_i^\textrm{X}q_k^\textrm{Y}),
\end{equation}
where $\textrm{X}$ and $\textrm{Y}$ can be either $\textrm{S}$ or 
$\textrm{I}$. In the definition, $p_{i,k}^\textrm{SS}$ denotes 
the probability that an SS link are attached by an $i$-degree S 
node and a $k$-degree S node, and $q_i^\textrm{S}$ is defined as 
$q_i^\textrm{S}=\sum_k{p_{i,k}^\textrm{SS}}$ denoting the edge 
end distribution of all the S nodes reached by the SS links. 
Hence, we have $\sum_{i,k}{p_{i,k}^\textrm{SS}=1}$, and 
$\sum_i{q_i^\textrm{S}}=1$, and 
$\sigma_e^\textrm{S}=\left({\sum_i{i^2q_i^\textrm{S}}-(\sum_i{iq_i^\textrm{S}})^2}\right)^{1/2}$. 
Moreover, $p_{i,k}^\textrm{II}$, $p_{i,k}^\textrm{IS}$, 
$q_i^\textrm{I}$ and $\sigma_e^\textrm{I}$ are similarly defined. 
For the initial condition, suppose that a fraction $\varepsilon$ 
of nodes are randomly selected as infected nodes. Then, the 
number of SS links whose ends are attached by an $i$-degree S 
node and a $k$-degree S node can be calculated as 
$(1-\varepsilon)(1-\varepsilon)p_{i,k}M$, and the total number of 
SS links is 
$(1-\varepsilon)(1-\varepsilon)\sum_{i,k}{p_{i,k}}M=(1-\varepsilon)(1-\varepsilon)M$ 
with $M$ being the total number of the connections. Therefore, we 
have 
$p_{i,k}^\textrm{SS}/\sum_{i,k}p_{i,k}^\textrm{SS}=p_{i,k}/\sum_{i,k}p_{i,k}$ 
and consequently $r_\textrm{SS}(0)=r(0)$. With similar analysis, 
we obtain 
$r(0)=r_\textrm{SS}(0)=r_\textrm{II}(0)=r_\textrm{IS}(0)$.

Figure \ref{Fig:Correlated-rSSrIIrIS} shows the evolution of 
these measures for the cases studied in 
Fig.\,\ref{Fig:Correlated-ir}. We can observe that, in all cases, 
$r_\textrm{SS}$ goes to $0$ most quickly as compared with other 
measures and $r_\textrm{II}$ have a significant drop in the 
beginning stage of the evolution. Besides, the speed of 
$r_\textrm{IS}$ moving towards $0$ is between those of 
$r_\textrm{SS}$ and $r_\textrm{II}$. This phenomenon can be 
understood as follows: for the susceptible nodes, those have 
larger degree can be infected more easily as they have more 
infected neighbors on average. Thus, the number of large-degree S 
nodes reduces quickly, which makes the degrees of the nodes at 
the ends of SS links quickly tend to small value. As a 
small-degree range confines the degree correlation at a low 
level, the value of $r_\textrm{SS}$ reduces accordingly. On the 
other hand, since in the beginning stage, large-degree 
susceptible nodes are easily infected, which contributes to the 
increase of the number of large-degree infected nodes and these 
large-degree infected nodes probably have infected neighbors with 
smaller degree, they could form II links with a negative 
correlation.

To verify our explanation, we plot the ratios of 
$p_{i,k}^\textrm{SS}$ and $p_{i,k}^\textrm{II}$ at $t=60$ to that 
at $t=0$ for the case of $r(0)=0.3$ in 
Fig.\,\ref{Fig:Correlated-GrayPlot}\,(a) and (c) and for the case 
of $r(0)=-0.3$ in Fig.\,\ref{Fig:Correlated-GrayPlot}\,(b) and 
(d), respectively. We observe that for both cases SS links mainly 
converge to the point (2,2) on the degree plane with $2$ being 
the average degree in this case at $t=60$, while 
$p_{i,k}^\textrm{II}$ increases in the large-degree range. In 
other words, for the case of $r(0)=-0.3$, the ratio of 
$p_{i,k}^\textrm{II}$ has an obvious increase in the negative 
correlation region.

\section{Conclusion}
\label{sec:Conclusion}

We have introduced a link-based formalism to present the adaptive
network model proposed by Gross et al. \cite{Gross:2006}. In our
method, inspired by the node classification method 
\cite{Marceau:2010} and master equations approach 
\cite{Gleeson:2011}, we took link as the object and classify them 
according to the disease states, the number of neighbors, and the 
number of infected neighbors of each node at its ends. Moreover, 
for clarity, we have separated the whole dynamics into three 
sub-processes, namely \emph{Recovery Process}, \emph{Infection 
Process} and \emph{Rewiring Process}. For each process, a set of 
differential equations is established to describe the evolution 
of each class of links. By combining all these equations 
together, an integrated formalism describing the time evolution 
of the adaptive networks is established.

Our formalism provides subtle detailed information of degree
correlation. It can predict the evolution of the system with 
higher accuracy than former node-based method where the network 
was treated under zero degree correlation approximation. More 
importantly, it can offer a more precise description of the 
network structure. Since the network topology is one of the key 
ingredients of an adaptive network, in-depth observation of the 
topology is of importance to reveal how the network topology and 
the nodal dynamics coevolve. Thus, our theoretical formalism 
could be a valuable improvement in understanding the mechanism of 
adaptive networks.

To reveal further details than those by the degree correlation 
measure $r$ \cite{Newman:2002}, we have introduced other degree 
correlation measures $r_\textrm{SS}$, $r_\textrm{II}$, 
$r_\textrm{IS}$ according to the disease states of the nodes at 
the ends of the links. We found that although the time evolution
of $r$ is relatively steady, the evolutions of $r_\textrm{SS}$, 
$r_\textrm{II}$ are not. Specifically, $r_\textrm{SS}$ moves to 
zero most quickly among all the measurements, while 
$r_\textrm{II}$ has a significant drop in the beginning stage. An 
explanation of this phenomenon can be provided with our formalism.

In this work, we assume that when susceptible nodes are rewiring 
their broken links, they do \emph{not} consider the nodal degree 
of the susceptible nodes to be selected. Thus, the system will 
tend to zero (or a near zero) correlation when the evolving time 
is long enough. However, in a more general situation, nodes may 
consider the nodal degrees of each other, and hence the system 
could possess non-zero correlation all the time. For this more 
general situation, our link-based method might be extended to 
handle the degree correlation in predicting the dynamics of such 
systems, and this will be a future research topic.

\begin{figure}[b]
\includegraphics[width=1.0\linewidth]{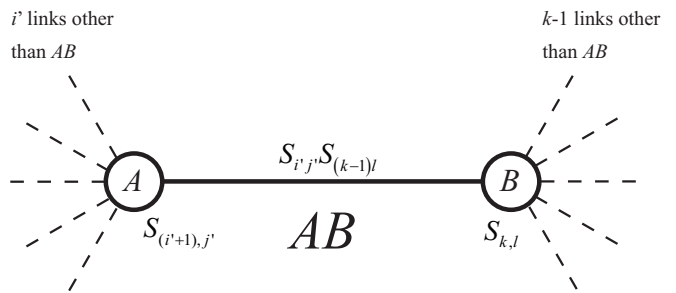}
\caption{\label{Fig:Deviation-Skl}Diagram of an $S_{k,l}$ node,
denoted as $B$, connected with an $S_{(i'+1),j'}$ node, denoted
as $A$, through an $S_{i'j'}S_{(k-1)l}$ link, denoted as $AB$.}
\end{figure}

\begin{figure*}
\includegraphics[width=1.0\linewidth]{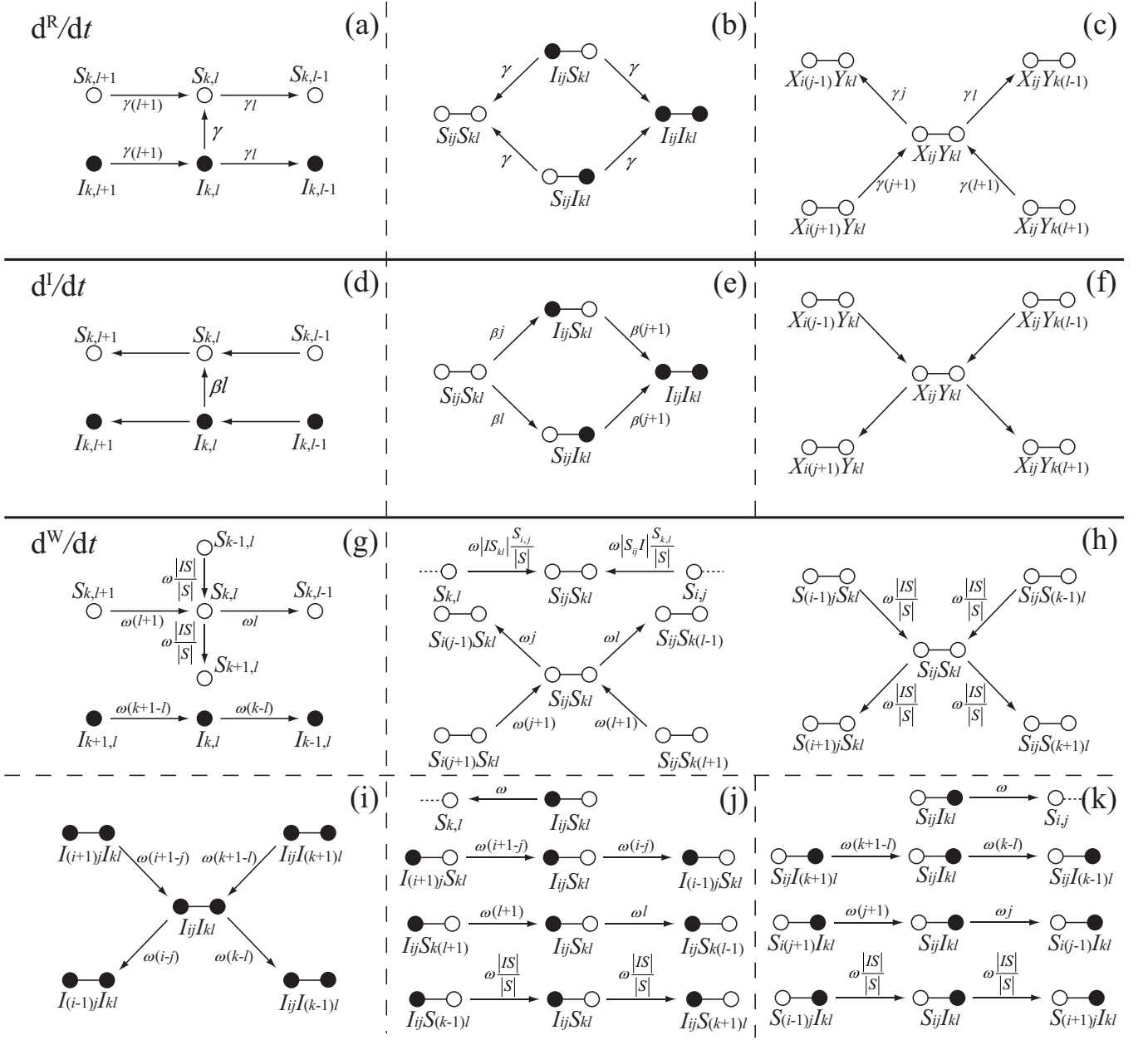}
\caption{\label{Fig:Diagram} Diagram of the variations of 
node/link class sizes in the Recovery Process 
$\textrm{d}^\textrm{R}/\textrm{d}t$ (a) - (c), the Infection 
Process $\textrm{d}^\textrm{I}/\textrm{d}t$ (d) - (f), and the 
Rewiring Process $\textrm{d}^\textrm{W}/\textrm{d}t$ (g) - (k). 
Open symbols ($\circ$) represent the susceptible nodes and solid 
symbols ($\bullet$) represent the infected nodes. $A \rightarrow 
B$ means the state of node or link represented by $A$ changes to 
that represented by $B$, and the term near an arrow 
'$\rightarrow$' indicates the rate of the corresponding change. 
Specifically, panel (a) [(d)] indicates the variations of the 
node states in the Recovery [Infection] Process, panel (b) [(e)] 
indicates the variations of the link states caused by the state 
change of the nodes at their ends, and panel (c) [(f)] indicates 
the variations of the links states caused by the state change of 
the neighbors of the their end nodes. In panels (c) and (f), $X$ 
and $Y$ can be either $S$ or $I$. In panel (f) the rates are 
complicated therefore are not shown here; details can be found in 
the text. Panel (g) indicates the variations of the node states 
in the Rewiring Process. Panel (h), (i), (j), (k) indicate the 
variation of the link states for the SS link, II link, IS link 
and SI link in the Rewiring Process, respectively, where 
$|IS|=\sum_{i,j,k,l}I_{ij}S_{kl}$, 
$|IS_{kl}|=\sum_{i,j}I_{ij}S_{kl}$, 
$|S_{ij}I|=\sum_{k,l}S_{ij}I_{kl}$, and $|S|=\sum_{k,l}S_{k,l}$.}
\end{figure*}

\appendix
\chapter{}
\section{Derivation of the second term on the right hand side of Eq.\,(\ref{eq:Infection-Skl})}
\label{App:Deviation-dIdtSkl}

This term is $-\beta\sum_{i',j'}{j'S_{i'j'}S_{(k-1)l}\langle k 
\rangle}$, which denotes the rate of the process that $S_{k,l}$ 
nodes change to $S_{k,l+1}$. This process happens when one of the 
susceptible neighbors of an $S_{k,l}$ node is infected.

As shown in Fig.\,\ref{Fig:Deviation-Skl}, we use $B$ to denote 
an $S_{k,l}$ node. Suppose that node $B$ has a neighbor $A$ which 
is an $S_{(i'+1),j'}$ node. Then, the link, denoted as $AB$, 
connecting $A$ and $B$, is an $S_{i'j'}S_{(k-1)l}$ link. Here, 
$k-1$ comes when we count the number of the neighbors of node $B$ 
we exclude the present link $AB$. Moreover, since node $A$ is in 
S state, we could infer that among the other $k-1$ neighbors of 
node $B$, $l$ of them are infected. Therefore, the fourth 
subscript of the link class that $AB$ belongs to is $l$. The 
meaning of $i'$ and $j'$ could be understood similarly. Since 
node $A$ has $j'$ infected neighbors, the probability that it is 
infected in one time step equals $\beta j'$.

Now, the question is: what is the probability that the node $A$ is
an $S_{(i'+1),j'}$ node? This is equivalent to ask: what is the
probability that a link connecting node $B$ is an
$S_{i'j'}S_{(k-1)l}$ link. Here, we use the mean-field theory to
estimate it. The total number of the links that have an $S_{k,l}$
node on one end and another S state node on the other end equals 
$(k-l)S_{k,l}N$. Among them, $S_{i'j'}S_{(k-1)l}\langle k \rangle 
N$ links satisfy the condition that one end is an $S_{k,l}$ node 
and the other end is an $S_{(i'+1),j'}$ node. Therefore, on 
average the probability that a link connecting node $B$ is an 
$S_{i'j'}S_{(k-1)l}$ link equals $[S_{i'j'}S_{(k-1)l}\langle k 
\rangle N]/[(k-1)S_{k,l}N]$, and this probability is also the one 
that node $A$ is an $S_{(i'+1),j'}$ node.

Since an $S_{(i'+1),j'}$ node has $j'$ infected neighbors, the
probability that node $A$ is an $S_{(i'+1),j'}$ node and gets
infection in one time step equals $[\beta 
j'S_{i'j'}S_{(k-1)l}\langle k \rangle]/[(k-l)S_{k,l}]$.

Furthermore, by summing up all the possible node classes that 
node $A$ may belong to, we obtain the probability that a 
susceptible neighbor of $B$ gets infection in one time step, as 
$[\beta\sum_{i',j'}j'S_{i'j'}S_{(k-1)l}\langle k 
\rangle]/[(k-l)S_{k,l}]$.

Since node $B$ has $(k-l)$ links connecting to its S state 
neighbors, the rate that $S_{k,l}$ nodes change to $S_{k,l+1}$ 
equals $(k-l)S_{k,l}[\beta\sum_{i',j'}j'S_{i'j'}S_{(k-1)l}\langle 
k \rangle]/[(k-l)S_{k,l}]$. Finally, since this process leads to 
the decrease of the fraction of $S_{k,l}$ nodes, we come to the 
result $-\beta\sum_{i',j'}{j'S_{i'j'}S_{(k-1)l}\langle k 
\rangle}$.

\chapter{}
\section{Derivation of the fifth term on the right hand side of Eq.\,(\ref{eq:Rewiring-SijSkl})}
\label{App:Deviation-dWdtSijSkl}

This term is $-\omega|IS|S_{ij}S_{kl}\langle k \rangle/|S|$, 
which denotes the rate of the process that links $S_{ij}S_{kl}$ 
change to $S_{(i+1)j}S_{kl}$. This process happens when an 
$S_{(i+1),j}$ node at one end of an $S_{ij}S_{kl}$ link is chosen 
in rewiring to form a new link.

First, the total amount of broken links is $\omega|IS|$ with 
$|IS|=\sum_{i,j,k,l}{I_{ij}S_{kl}}$ being the total amount of IS 
link. Then, for each broken link, the probability that it chooses 
an $S_{(i+1),j}$ node equals $S_{(i+1),j}/|S|$, where 
$|S|=\sum_{k,l}S_{k,l}$. When an $S_{(i+1),j}$ node is chosen by 
a rewired link, the states of the links attached on this node 
will also be changed. By adopting the mean-filed theory 
approximation, for a link attached on an $S_{(i+1),j}$ node, the 
probability that it is an $S_{ij}S_{kl}$ link equals 
$[S_{ij}S_{kl}\langle k \rangle N]/[(i-j+1)S_{i+1,j}N]$, where 
the denominator represents the total number of the links of which 
one end is an $S_{(i+1),j}$ node and the other end is a 
susceptible node, while the numerator represents the number of 
$S_{ij}S_{kl}$ links. Since an $S_{(i+1),j}$ node has $(i-j+1)$ 
such links, the average number of the $S_{ij}S_{kl}$ links that 
an $S_{(i+1),j}$ node has is equal to $\{(S_{ij}S_{kl}\langle k 
\rangle N)/[(i-j+1)S_{i+1,j}N]\}(i-j+1)$.

Summing up all the above, the portion that the $S_{ij}S_{kl}$ 
links changing to $S_{(i+1)j}S_{kl}$ links in this process equals 
$\omega|IS|\{S_{i+1,j}/|S|\}\{S_{ij}S_{kl}\langle k \rangle 
N/[(i-j+1)S_{i+1,j}N]\}(i-j+1)$. Since this process leads to the 
decrease of the fraction of $S_{ij}S_{kl}$ links, we come to the 
conclusion that the item is $-\omega|IS|(S_{ij}S_{kl}\langle k 
\rangle)/|S|$.

\chapter{}
\section{A comprehensive diagram of all the processes described in the ODEs}
\label{App:Diagram}

This diagram is shown in Fig.\,\ref{Fig:Diagram}.

\acknowledgments

This work is supported by the Ministry of Education, Singapore, 
under contract RG69/12; the Hong Kong Research Grants Council 
under the GRF Grant CityU 1109/12E; and the European Commission 
through FET-Proactive project PLEXMATH (FP7-ICT-2011-8; grant 
number 317614). We acknowledge James P. Gleeson for beneficial 
comments and discussions.


\begin{references}
\bibitem{Notation}
As to the notations of the classes, for node classes the pair of 
subscripts are separated with a comma (e.g. $S_{k,l}$); for link 
classes the two pairs of subscripts are not separated, as 
a distinction to node classes (e.g. $S_{ij}S_{kl}$). When a 
subscript of a link class has ``$+1$" or ``$-1$", we put it 
between a pair of parentheses for indication (e.g. 
$S_{i(j+1)}S_{(k-1)l}$). However, parentheses are not used for 
node classes when a subscript has ``$+1$" or ``$-1$" (e.g. 
$S_{k+1,l}$).
\end{references}

\end{document}